%% file: 00-main.tex
\documentclass{article}

\usepackage{microtype}
\usepackage{graphicx}
\usepackage{subfigure}
\usepackage{booktabs} 

\usepackage{hyperref}

\usepackage{tikz}
\usepackage{mdframed}
\usepackage[normalem]{ulem}
\usepackage{enumitem}
\usepackage[skip=0pt]{caption}
\usepackage{balance}
\usepackage{xspace}

\usepackage[accepted]{mlsys2024}

\input{macros}

\mlsystitlerunning{Just-in-time Quantization with Processing-In-Memory for Efficient ML Training}

\begin{document}

\twocolumn[
\mlsystitle{Just-in-time Quantization with \\ Processing-In-Memory for Efficient ML Training}



\mlsyssetsymbol{equal}{*}

\begin{mlsysauthorlist}
\mlsysauthor{Mohamed Assem Ibrahim}{amd}
\mlsysauthor{Shaizeen Aga}{amd}
\mlsysauthor{Ada Li}{amd}
\mlsysauthor{Suchita Pati}{amd}
\mlsysauthor{Mahzabeen Islam}{amd}
\end{mlsysauthorlist}

\mlsysaffiliation{amd}{Advanced Micro Devices, Inc.}

\mlsyscorrespondingauthor{Mohamed Assem Ibrahim}{mohamed1.ibrahim@amd.com}

\mlsyskeywords{LLM training, Processing-in-memory, Quantization}

\vskip 0.3in
\vspace{-0.1\baselineskip}

\begin{abstract}
\input{01-abstract}
\end{abstract}
]



\printAffiliationsAndNotice{}  

\input{02-introduction}
\input{03-background}
\input{04-motivation} 
\input{05-jit-q-on-pim}
\input{06-evaluation}
\input{07-discussion}
\input{08-related}
\input{9-conclusions}

\section*{Acknowledgements}
AMD, the AMD Arrow logo, and combinations thereof are trademarks of Advanced Micro Devices, Inc.  Other product names used in this publication are for identification purposes only and may be trademarks of their respective companies.


\balance
\bibliography{ref}
\bibliographystyle{mlsys2024}



\end{document}

%% file: macros.tex
\long\def\ignore#1{}
\sloppypar

\newcommand*\circled[1]{\tikz[baseline=(char.base)]{
  \node[shape=circle,draw,fill=black,text=white,font=\bf,inner sep=0.5pt] (char)
  {\scriptsize#1};
}}

\newcommand{\ApproxSign}{\raise.17ex\hbox{$\scriptstyle\sim$}}

\newcommand{\putsec}[2]{\vspace{-0.0in}\section{#2}\label{sec:#1}\vspace{-0.0in}}
\newcommand{\putssec}[2]{\vspace{-0.0in}\subsection{#2}\label{ssec:#1}\vspace{-0.0in}}
\newcommand{\putsssec}[2]{\vspace{-0.0in}\subsubsection{#2}\label{sssec:#1}\vspace{-0.0in}}

\newcommand{\tabref}[1]{Table~\ref{#1}}
\newcommand{\figref}[1]{Figure~\ref{#1}}
\newcommand{\secref}[1]{Section~\ref{sec:#1}}
\newcommand{\ssecref}[1]{Section~\ref{ssec:#1}}
\newcommand{\sssecref}[1]{Section~\ref{sssec:#1}}

\mdfdefinestyle{verdictbox}
{
    linewidth=0.7pt,
    innertopmargin=2pt,
    innerbottommargin=2pt,
    innerrightmargin=2pt,
    innerleftmargin=2pt,
    leftmargin=0pt,
    rightmargin=0pt,
    nobreak=false,
    splitbottomskip=0pt,
}

\newcommand{\squishlist}{
   \begin{list}{$\bullet$}
    { \setlength{\itemsep}{0pt}      \setlength{\parsep}{0pt}
      \setlength{\topsep}{3pt}       \setlength{\partopsep}{0pt}
      \setlength{\listparindent}{-2pt}
      \setlength{\itemindent}{-5pt}
      \setlength{\leftmargin}{1em} \setlength{\labelwidth}{0em}
      \setlength{\labelsep}{0.5em} } }

\newcommand{\squishend}{
    \end{list}  }

\newcommand{\squishlisttwo}{
   \begin{list}{$\bullet$}
    { \setlength{\itemsep}{0pt}    \setlength{\parsep}{0pt}
      \setlength{\topsep}{0pt}     \setlength{\partopsep}{0pt}
      \setlength{\leftmargin}{2em} \setlength{\labelwidth}{1.5em}
      \setlength{\labelsep}{0.5em} } }

%% file: 01-abstract.tex
Data format innovations have been critical for machine learning (ML) scaling, which in turn fuels ground-breaking ML capabilities. 
However, even in the presence of low-precision formats, model weights are often stored in both high-precision and low-precision during training. 
Furthermore, with emerging directional data formats (e.g., MX9, MX6, etc.) multiple low-precision weight copies can be required. 
To lower memory capacity needs of weights, we explore just-in-time quantization (JIT-Q) where we only store high-precision weights in memory and generate low-precision weights only when needed. 
To perform JIT-Q efficiently, in this work, we evaluate emerging processing-in-memory (PIM) technology to execute quantization. 
With PIM, we can offload quantization to in-memory compute units enabling quantization to be performed without incurring costly data movement while allowing quantization to be concurrent with accelerator computation. 
Our proposed PIM-offloaded quantization keeps up with GPU compute and delivers considerable capacity savings (up to 24\%) at marginal throughput loss (up to 2.4\%). 
Said memory capacity savings can unlock several benefits such as fitting larger model in the same system, reducing model parallelism requirement, and improving overall ML training efficiency.

%% file: 02-introduction.tex
\putsec{s01}{Introduction}

Model scaling has been a critical component to unlocking disruptive machine learning (ML) capabilities. 
This scaling, however, has not been matched with commensurate memory capacity scaling~\cite{ZeRO-Infinity21}. 
These conflicting trends have led to lower efficiency for ML due to increased reliance on distributed computing (communication overheads), lower batch-sizes (lower compute efficiency) and more. 
As such, techniques which optimize memory capacity needs of ML stand to lower these overheads and can lead to increased ML efficiency. 

To tackle the memory capacity challenge, there has been considerable interest in quantization of tensors to (increasingly) low-precision numeric formats such as BF16~\cite{Kalamkar19BF16}, FP8~\cite{micikevicius2022fp8}, and beyond. 
By storing and computing on tensors in low-precision formats instead of single-precision (FP32) format, considerable capacity savings and also compute efficiency can be attained. 
As such, data formats continue to be an active area of investigation, with emerging shared microexponents (MX) formats~\cite{Rouhani23MX} further pushing the precision down to four bits. 

\begin{figure}[t]
    \centering
    \includegraphics[width=\linewidth]{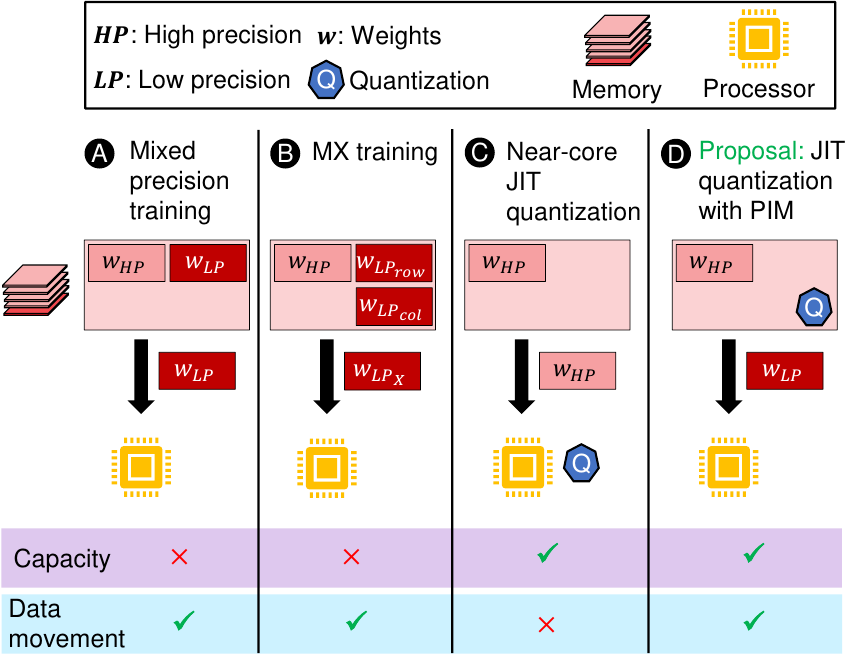}
    \caption{JIT-Q achieves capacity savings while maintaining the data movement savings.}
    \label{fig:intro_compare}
    \vspace{-\baselineskip}
\end{figure}

While this data format evolution has been a key lever in optimizing ML capacity needs, we believe there exist data redundancy in ML training which can be optimized for capacity savings. 
Specifically, state-of-art training techniques which employ low-precision formats, typically employ mixed-precision training~\cite{Micikevicius17}, wherein to attain good accuracy, weight tensors are maintained in both high-precision and low-precision in memory. 
The high-precision copy of weights accumulates the gradients after each optimizer step. 
At the same time, a quantized copy of these high-precision weights, in low-precision, is maintained, which is employed in computations for forward and back-propagation (\figref{fig:intro_compare}~\circled{A}) phases. 
Further, with directional (MX) data formats, which require quantization to be applied along the reduction dimension, two low-precision copies of weights, to be used in forward and back-propagation respectively, can be necessary (\figref{fig:intro_compare}~\circled{B}). 

An observation we make in this work is that as the high-precision copy of weights is necessary for training, and the low-precision copy is derived from the high-precision copy, a mechanism to cheaply create a low-precision copy of weights, as and when needed, can obviate the need to store the low-precision weight copy in memory and thus save memory capacity. 
We term this \textit{just-in-time} quantization (JIT-Q). 
Note that, as in \figref{fig:intro_compare}~\circled{C}, while a straightforward way to accomplish this is to read-in the high-precision weight copy for computations and quantize the weights before use at the core, this causes increased (and unnecessary) data movement, which is the chief contributor to ML energy expenditure~\cite{untetherai_datamov}.
Further, this also places quantization on the critical path. 
While this can be mitigated by scheduling quantization kernels ahead of time and in concurrence with main GPU computation, doing so causes even more data movement as not only will the high-precision copy be read, the low-precision copy will also be written. 

Instead, in this work, we propose to harness emerging processing-in-memory (PIM) technology to perform the aforementioned JIT-Q of weight tensors. 
With recent functional PIM prototypes from multiple memory vendors~\cite{mi100PIM,hynixPIM}, commands are broadcasted to in-memory compute units and data is operated in-place in memory instead of moving data to the accelerator (e.g., GPU). 
While the accelerator coupled with memory can access one memory bank at a time over a shared data bus, by not using the shared data bus, PIM enables data from multiple banks to be operated on in tandem. 
This provides considerable memory bandwidth boost over that available to accelerator all the while enabling computation over data without incurring costly data movement. 
In this work we focus on high-bandwidth memory (HBM) PIM, as HBM is coupled with GPUs, the most ubiquitous ML accelerators.

As depicted in \figref{fig:intro_compare}~\circled{D}, PIM enables harnessing capacity savings of JIT-Q of weights without incurring costly data movement. 
To do so, we first deduce a quantization routine that can be offloaded to in-memory compute units in HBM (we evaluate HBM-PIM architecture as discussed in~\cite{samsungPIM} in this work). 
We identify data placement considerations (for both scalar formats and block-based directional formats) and in-memory ALU augmentations that lead to an efficient quantization routine. 
Further, we discuss how this routine can be co-scheduled with main GPU computation to deliver quantized low-precision weight tensors just-in-time to the GPU computation. 

Our analysis across current and future models shows that our proposed PIM-offloaded quantization keeps up with main GPU computation and delivers considerable capacity savings (up to 24\%) at marginal throughput loss (up to 2.4\%). 
Resultant memory capacity savings unlock benefits that improve overall compute utilization; it enables devices to fit larger models and/or larger batch-sizes (which may help with faster convergence), and reduces reliance on model parallelism which requires additional communication.

Overall, our work makes the following key contributions:
\vspace{-\baselineskip}
\begin{itemize}[noitemsep,nolistsep,leftmargin=*]

    \item We propose just-in-time quantization (JIT-Q) for weights which enables storing of only high-precision copy of weights during training and creates low-precision weight copies just when they are needed. 
    
    \item We evaluate efficacy of emerging commercial processing-in-memory (PIM) solutions to perform JIT-Q. 
    Offloading to PIM enables quantization of weights without incurring costly data movement.
    
    \item We discuss data placement and in-memory ALU augmentations necessary to offload quantization to PIM efficiently. 
    Our evaluations show that our proposed PIM quantization routine can prepare low-precision weights copy just-in-time without stalling concurrent GPU computation. 
    
    \item Proposed JIT-Q of weights delivers memory capacity savings of up to 24\% which can be harnessed in many ways for efficient ML training.
    
\end{itemize}

%% file: 03-background.tex
\putsec{s02}{Background}

\putssec{bckg_llm}{Large Language Models}
We focus in this work on Transformer-based~\cite{vaswani2017attention} large language models (LLMs) given their applicability across domains and modalities~\cite{tsimpoukelli2021multimodal,sung2022vl,alayrac2022flamingo}.
From a computational perspective, LLMs have two training phases, an expensive, but one-time, pre-training phase for general learning and another short task-specific fine-tuning phase. Post learning, LLMs are deployed for inference. The basic building block of an LLM is an \textit{encoder} or a \textit{decoder} layer. These layers are made of a multi-head attention sub-layer and a multi-layer perceptron (MLP) sub-layer. Operations in these sub-layers manifest as matrix multiplication operations (GEMMs) followed by a few element-wise and reduction operations (e.g., residual connection, layer normalization) which are often fused with the GEMMs. The encoder and decoder layers are similar except the decoder's attention sub-layer/GEMM input is masked, which causes different inference behavior computationally but does not affect training which is the focus of this work. Both training phases involve forward and backward propagation through the layers, followed by parameter updates. In contrast, inference phase does not have backward propagation and weight update. 
Note, while we anchor on LLMs, ideas proposed in this work are applicable to other model architectures.

\putssec{bckg_mx}{Directional Blocked Data Formats}

Continued scaling of ML models has been a key ingredient to their disruptive capabilities. A critical fuel to this scaling is evolution of low-precision data formats. Low-precision formats reduce memory capacity requirement and consequently data movement along with delivering performance improvement by harnessing higher throughput compute. While an active area of research, in this paper, we focus on shared microexponents (MX) format~\cite{Rouhani23MX}, an emerging low-precision data format, because of the balance it achieves between maintaining model accuracy, improving hardware efficiency, while reducing software friction. We discuss other formats in \secref{s07}.

\begin{figure}[t]
    \centering
    \includegraphics[width=\linewidth]{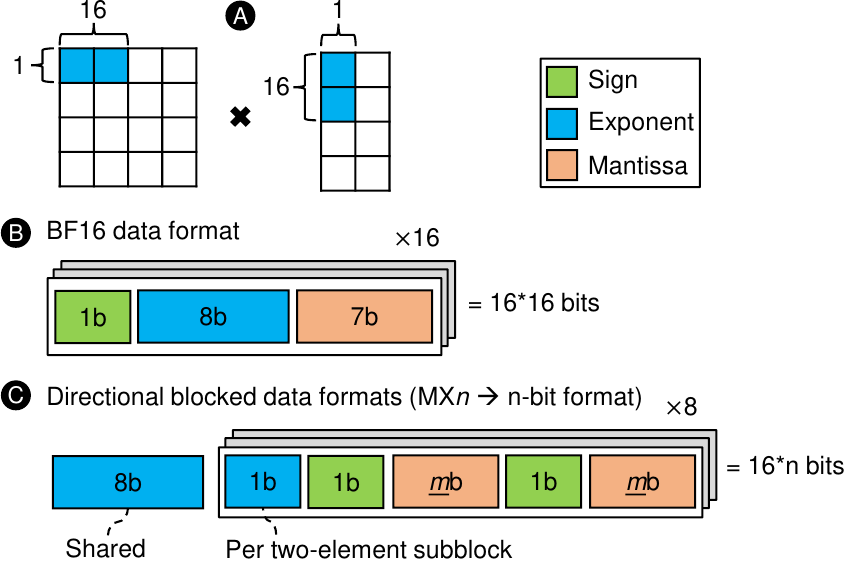}
    \caption{Directional data formats.}
    \label{fig:bg_data_format}
    \vspace{-1.5\baselineskip}
\end{figure}

As shown in \figref{fig:bg_data_format}, MX formats are directional blocked data formats that represent a block of $N$ elements ($N$=16 BF16 elements in the figure).  Instead of using a per-element sign, exponent, and mantissa bits as in scalar data formats~\circled{B}, MX formats settles for only sign and mantissa bits per element~\circled{C}. Based on the number of mantissa bits, there are three variants of MX formats to tailor to the different needs of training and inference. Specifically, MX9, MX6, and MX4 use 7, 4, and 2 mantissa bits per element, respectively. As for exponent bits, MX format uses two-levels of scaling factors (exponents) to reduce the negative effects of outliers and provides additional quantization noise reduction. That is, an 8b exponent is shared across \textit{all} $N$ elements, and a second-level 1b exponent is shared among a subblock of two elements. Finally, as shown in~\circled{A}, for a GEMM operation where both interacting tensors are MX-quantized, to harness benefits of MX formats, the tensors have to be quantized along the reduction dimension (blocking/quantization along row for one, blocking/quantization along column for other). 

\putssec{bckg_pim}{Commercial PIM Solutions}

\begin{figure}[t]
    \centering
    \includegraphics[width=\linewidth]{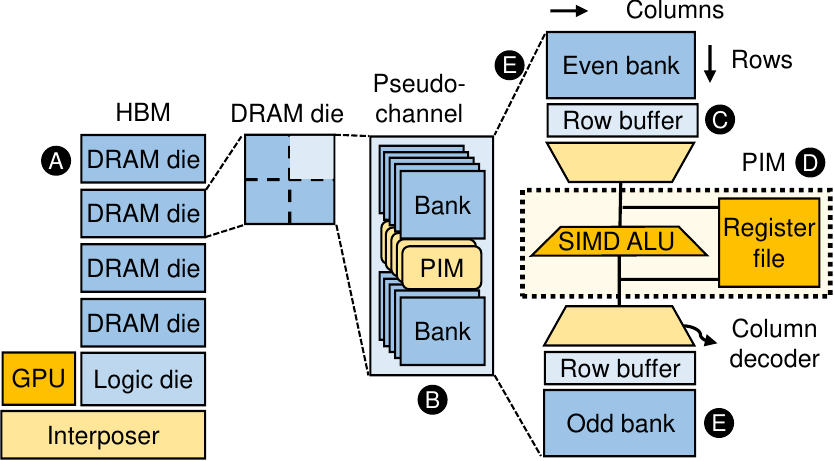}
    \caption{HBM PIM overview.}
    \label{fig:bg_hbm_dram_pim}
    \vspace{-1.5\baselineskip}
\end{figure}

To cater to increasing demands for memory bandwidth from emerging applications, memory vendors like Samsung and SK Hynix have proposed commercial processing-in-memory (PIM) solutions for DRAM based memories ~\cite{samsungPIM,kim2022aquabolt,hynixPIM,lee2019design,he2020newton}. 
With PIM, compute units are placed on the periphery of core DRAM structures, thus avoiding changes to internal DRAM structures and making commercialization easier. In a recent real PIM prototype ~\cite{samsungPIM}, Samsung has integrated PIM with high bandwidth memory (HBM), which is coupled with GPUs.
Figure~\ref{fig:bg_hbm_dram_pim} depicts the PIM design we focus on in this work.

HBM provides high bandwidth and energy efficiency via high density interconnects and in-package 2.5D integration with the processor~\cite{HBM-jedec}~\circled{A}. 
HBM memory access requires similar set of basic operations as conventional DRAM; however, HBM offers wider interface. 
Each HBM die comprises multiple pseudo-channels (pCHs)~\circled{B}. 
Each pCH contains multiple banks sharing the data bus of the pCH. 
Further, similar to generic DRAM, a bank is comprised of multiple rows and columns. 
On a memory access from the GPU, a row worth data is brought into the per-bank row-buffer~\circled{C} incurring row activation overhead) and from there on column worth data (or multiples of it) can be accessed over the data bus from the GPU.

\begin{figure*}[t]
    \centering
    \includegraphics[width=\linewidth]{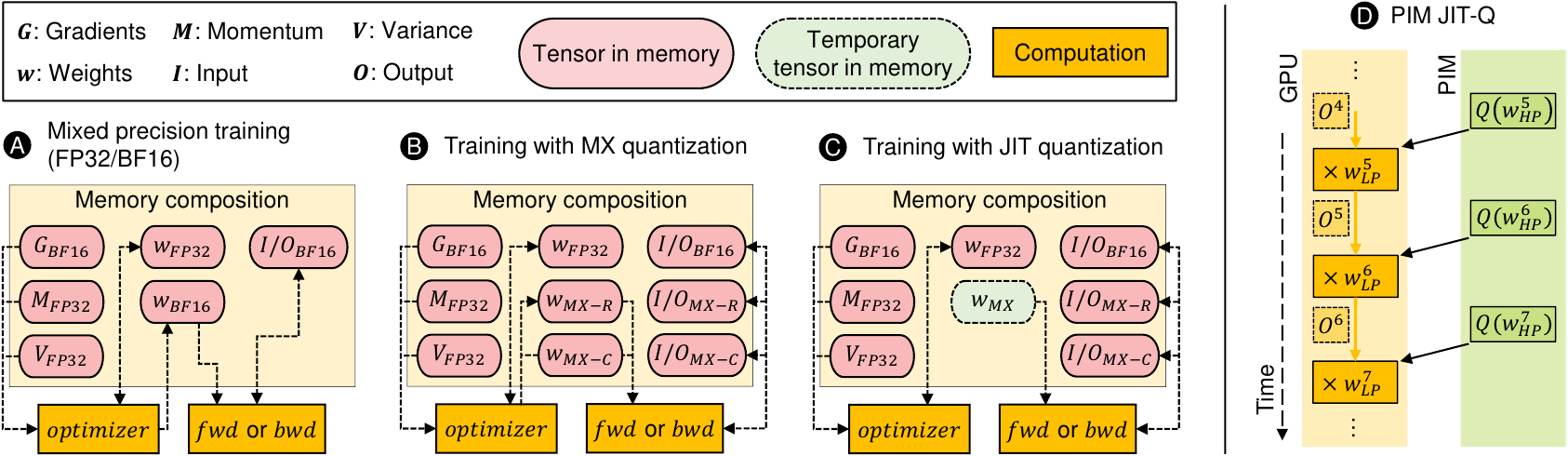}
    \caption{Memory composition, dataflow and redundancy in state-of-art training techniques \circled{A} Mixed-precision training and \circled{B} Training with MX-quantization. Proposed work optimizes weight redundancy with just-in-time quantization (JIT-Q) \circled{C}. Concurrent GPU and PIM execution (\circled{D}) for efficient JIT-Q of weights.}
    \label{fig:motiv_jit_q_train}
    \vspace{-\baselineskip}
\end{figure*}

In the HBM-PIM prototype~\cite{samsungPIM}, each PIM unit includes a 256b wide SIMD PIM ALU (processing on multiple lanes in parallel) and limited number of register files to temporarily store data~\circled{D}. 
To harness performance while managing area overheads, each PIM unit is shared by two banks (even and odd)~\circled{E}. 
To reduce complexity, the PIM units do not have any instruction fetch capabilities, rather PIM commands are sent by the GPU to the pCH. 
Each PIM command is broadcast to all PIM units inside the pCH and the PIM units operate in parallel. Therefore, PIM has bandwidth advantage over GPU (about 4-8$\times$) - GPU memory accesses to different banks of a pCH get serialized over the memory interface - in contrast all PIM units can independently access the attached banks. 
This bandwidth boost can be harnessed to offload bandwidth-intensive, low compute-to-byte, computations to PIM, while keeping compute-bound phases on GPU. 
Such collaboration is promising for workload acceleration~\cite{aga2019_coml}.

%% file: 04-motivation.tex
\putsec{s03}{Case for Just-in-time Quantization with PIM}

\putssec{llm_cap}{Capacity - A Key Performance Determinator}

Memory characteristics (capacity, bandwidth, latency) play a critical role in ML training efficiency. 
Specifically, in this work, we focus on capacity, which dictates the amount of data, and the associated computation, that gets mapped to an accelerator and as such computation efficiency. 
Model scaling in the recent past has surpassed capacity scaling, especially considering HBM capacity that is coupled with accelerators such as GPUs, which are commonly employed for ML training. 
This in turn has caused training state (optimizer state such as momentum, variance, gradient, weight tensors and intermediate state such as input, output tensors) to be sharded~\cite{narayanan2021efficient, zhao2023pytorch} across accelerators or offloaded away from HBM~\cite{ren2021zerooffload}. 
Such tensor sharding/offloading can have considerable impact on ML training efficiency. 
Specifically, such mechanisms reduce overall compute efficiency; communication overhead to gather/scatter tensors, smaller tensors due to sliced weights and lower batch.
Given its multi-dimensional impact, optimizing ML capacity needs is paramount to attaining better ML efficiency.

\putssec{llm_cap_redundancy}{Opportunity: Eliminating Weight Redundancy}
Prior works have optimized away memory redundancy in ML training to better support model scaling. 
As an example, works like ZeRo~\cite{samyam2020zero}, store a single copy of optimizer state partitioned across accelerators in a distributed training setup. 
While these prior works have considerably reduced memory redundancy, in this work, we observe that there is still further scope to reduce memory redundancy for ML training. 
Consider memory composition of an accelerator for mixed-precision training~\cite{Micikevicius17}, the defacto training technique, as depicted in \figref{fig:motiv_jit_q_train}~\circled{A}. 
With mixed-precision, weight tensors manifest redundancy as both high and low-precision copies of weights are stored in memory to be used in the optimizer computation and forward/back-propagation computations, respectively. 
This redundancy is worsened for training with directional blocked formats such as MX formats (described in \ssecref{bckg_mx}) for two variations of low-precision weights, quantized along different dimensions, are required (\figref{fig:motiv_jit_q_train}~\circled{B}). 
Note that, as high-precision weight tensors are necessary for effective training (to preserve small-valued updates),
a cheap mechanism which creates low-precision weight tensors only when needed can eliminate weight tensor redundancy and deliver capacity savings. 

\putssec{jit-q-pim-case}{PIM for Just-in-time Weight Quantization}
The above identified opportunity can be harnessed with just-in-time quantization (JIT-Q) of weight tensors. 
That is, by storing only high-precision weight tensor in memory and creating the low-precision weight tensor only when needed (\figref{fig:motiv_jit_q_train}~\circled{C}), weight tensor redundancy can be eliminated and capacity savings can be harnessed. 
However, to truly realize the benefits of this approach, an efficient mechanism to create low-precision weight tensor is necessary. 
To that end, note first that a naive mechanism which reads in high-precision weight tensor during forward and back-propagation computation causes unnecessary data movement and energy expenditure, and additionally adds quantization to critical path. 
While the latter of the problems can be tackled by co-scheduling quantization kernels ahead of time, such kernels, when executed by the accelerator will further add to data movement overheads by having to write-back low-precision weights to memory, as well as contending with the concurrent main computation for compute/memory resources. 

To tackle above challenges, in this work, we propose to harness processing-in-memory (PIM) technology and offload JIT-Q to in-memory compute units. 
With PIM, high-precision weight tensors are read by in-memory compute units, which quantize them to low-precision and store back the result to memory without incurring costly data movement from memory to the processor. 
By co-scheduling PIM computation with main acceleration computation (\figref{fig:motiv_jit_q_train}~\circled{D}), said low-precision tensors are only \textit{temporarily} created just when needed and discarded thereafter. 
Overall, this enables JIT-Q of weights to harness capacity savings while avoiding costly data movement. 
However, to do so, quantization should be efficiently offloaded to PIM  so as to keep up with main acceleration computation. 
We discuss techniques to do so next. 

%% file: 05-jit-q-on-pim.tex
\putsec{s05}{Efficient JIT-Q with PIM}

To efficiently offload a computation to PIM, a programmer first deduces a computation-conscious data mapping to exploit the strengths of the target PIM design (e.g., command broadcasts) while avoiding its shortcomings (e.g., no inter-bank communication, no cross SIMD compute). 
Next, a \textit{PIM kernel} which expresses the computation orchestration on the in-memory compute units is launched. 
In this section, we discuss the key considerations for offloading JIT-Q computation to PIM in terms of compute orchestration, data mapping, and augmentations to existing PIM design.

\putssec{jitq-pim-routine}{PIM Quantization Routine}

To initiate the required PIM computations on commercial PIM designs, host GPU launches \textit{PIM kernels}~\cite{samsungPIM,mi100PIM}.
These kernels are like existing GPU kernels except they issue \textit{pim-instructions}. A \textit{pim-instruction} effectively enqueues a \textit{pim-command} at the memory controller which in turn instructs PIM unit to execute either SIMD compute operation (e.g., add, multiply, etc.) or data movement (moving data between row-buffer and register file) along with necessary row activation. Finally, to feed the independent memory channels in GPUs, different workgroups (groups of threads or thread blocks), within the PIM kernel, issue PIM commands to different channels, which in turn are broadcasted to banks within a channel.
In addition to PIM ALU functionality supported in current PIM prototypes~\cite{samsungPIM}, to effectively support quantization, we assume the PIM ALU to support two operations: compare two operands (\textit{pim-CMP}) and single-bit intra-SIMD lane shift operation (\textit{pim-bitSHIFT}). 
Given their simplicity and support for related functionality in existing PIM prototypes, such as activation functions in Hynix-PIM~\cite{hynixPIM} and ReLU in Samsung-PIM~\cite{samsungPIM}, we believe this to be a reasonable assumption.

\begin{figure}[t]
    \centering
    \includegraphics[width=\linewidth]{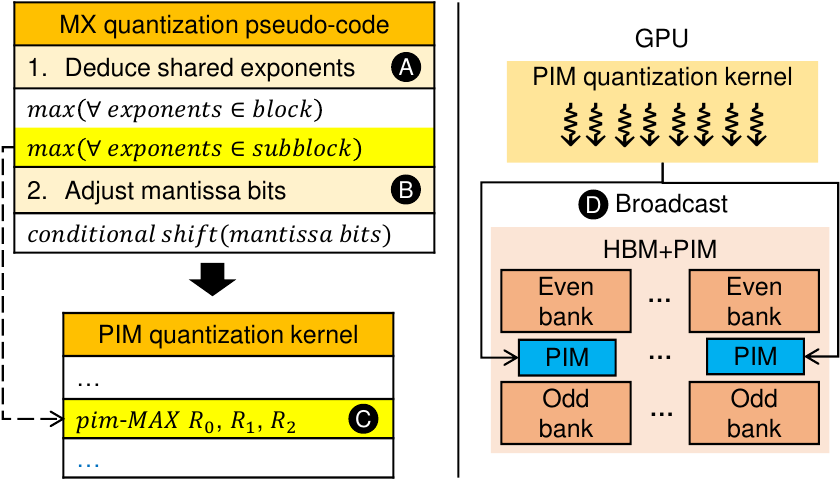}
    \caption{MX quantization pseudo-code and resultant PIM quantization kernel (left). PIM quantization kernel orchestration (right).}
    \label{fig:jit_q_pim_routine}
    \vspace{-\baselineskip}
\end{figure}

We describe this PIM routine for the MX format (\ssecref{bckg_mx} as tackling quantization for them is considerably more involved than for scalar formats. For MX quantization from scalar data formats, the input tensor is broken into blocks of $N$ elements ($N$=16). Then, for each $N$-element block, two key steps are performed as shown in \figref{fig:jit_q_pim_routine}. First, the quantization routine computes the shared level-1 exponent using a reduction function (e.g., max) of all exponents in the $N$ input elements~\circled{A}. Then, using this level-1 exponent, level-2 exponents are deduced for every  two input element block. Second, the per-element mantissa bits are adjusted, using bit-level shift operations, to compute the $m$-bit mantissa per output element~\circled{B}. We translate this pseudo-code to PIM quantization kernel via deducing the PIM instructions necessary to realize the computation. As an example, for exponent calculation, a series of \textit{pim-MAX} commands~\circled{C} are used, while for mantissa adjustment \textit{pim-bitSHIFT} commands are used. Using our PIM kernel, we broadcast these commands to banks in a channel (and to multiple channels) to fully exploit PIM parallelism~\circled{D}.

\noindent\textbf{Optimization.}
In contrast to scalar format quantization, the shift amount can differ for each input element as it depends on the level-2 exponent and the per input element's original exponent.
To efficiently support the different bit-level shift amounts, we augment the SIMD PIM ALU with counter-based conditional intra-lane shifts. 
This stores the per-lane shift amount in a register, then per each \textit{pim-bitSHIFT} command, it checks the stored shift amount per lane ($S_i$). 
If $S_i > 0$, then the shift is processed for the corresponding lane, then $S_i$ is decremented. Otherwise, the shift is skipped for the corresponding lane. 
Without such support, three PIM commands (\textit{pim-CMP}, \textit{pim-bitSHIFT}, and \textit{pim-ADD}) are used for a single bit shift, which increases the PIM quantization time (evaluated in \secref{s06}).

\putssec{jitq-pim-data-placement}{Weight Tensor Placement}

\begin{figure*}[t]
    \centering
    \includegraphics[width=\linewidth]{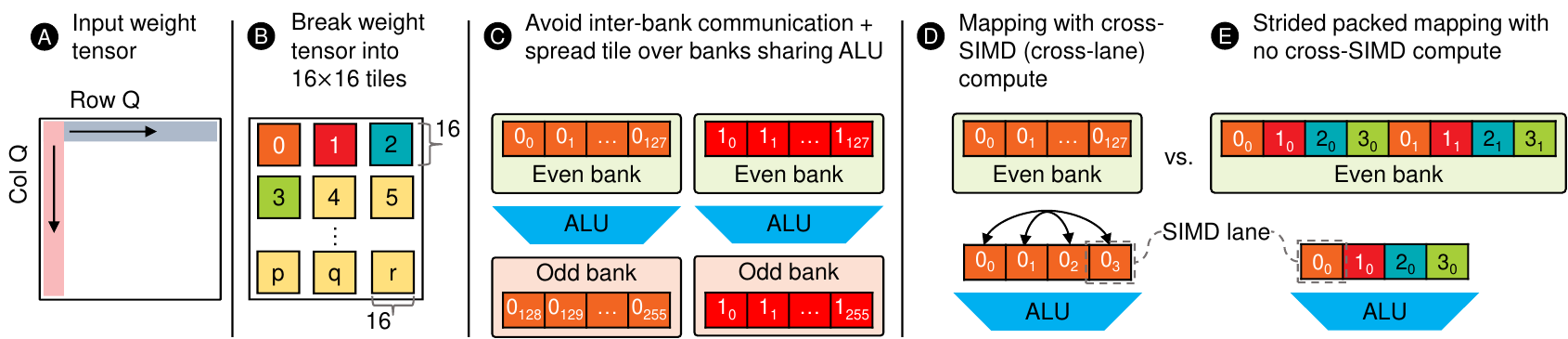}
    \caption{Placement of weight tensor in memory for efficient offload of MX quantization to PIM.}
    \label{fig:jit_q_mapping_strided_packing}
    \vspace{-\baselineskip}
\end{figure*}

Placement of data in memory is an important consideration for efficient computation offload to PIM for several reasons. First, our evaluated PIM design~\cite{samsungPIM} places a PIM compute unit, with a SIMD ALU, per two DRAM banks. Therefore, unless any interacting elements in the offloaded computation are mapped to the same bank (or banks sharing a PIM unit), inter-bank communication overhead is incurred. 
With the absence of direct inter-bank communication in current commercial PIM designs, the GPU performs this communication by copying data from one bank to another, which is expensive. 
Second, PIM broadcasts the same command to multiple banks in the same pseudo-channel to achieve its significant memory bandwidth boost. Therefore, proper interleaving of input/output data across banks/channels is required to harness the broadcast feature of PIM. Finally, while the SIMD ALU in the PIM compute unit helps harness data parallelism, it lacks support for cross-lane computations, resulting in lane-level shift operations. 
With the limited number of metal layers in the current DRAM technology, supporting different levels of bit shifts (single-bit shift for mantissa bits processing and lane-level shift for cross-lane computations) is likely to be costly. 
Given these considerations, we discuss our data mapping tailored to JIT-Q of MX formats as shown in \figref{fig:jit_q_mapping_strided_packing}.  

\noindent\textbf{Avoiding Inter-bank Communication.}
First, as MX formats require the input weight tensor to be quantized along the reduction dimension, supporting both row and column quantization is required as discussed in \ssecref{bckg_mx}. Therefore, the overall quantization computation involves elements along both row and column dimensions~\circled{A}. A naive MX-oblivious row-major mapping of the input tensor that divides the elements among the available PIM ALUs, to exploit the inherent parallelism of PIM, would map the row elements to same bank (or banks sharing a PIM compute unit) but not the column elements thus, triggering inter-bank communication during column quantization. To avoid this, we propose a tiled data mapping in which we break the input tensor into $N\times N$ tiles~\circled{B}, where $N$ is number of element per MX block, and map each 2D tile to a single bank~\circled{C}. 
The per-tile elements are mapped in a row-major fashion.
We term this tiled mapping as \textit{pim-jitq-tiled}.
Using row-major mapping within a tile is beneficial for row quantization as it minimizes the number of occupied DRAM rows per MX block, reducing the row activations overhead.
In contrast, it results in higher row activations overhead for the column quantization.
To reduce this overhead, we exploit the sharing of the PIM ALU between pair of DRAM banks and propose to spread the input tile over the even and odd banks~\circled{C}.

\noindent\textbf{Harnessing PIM Parallelism.}
To unlock PIM's full potential, we process multiple independent tiles in different PIM units via command broadcasting. 
The large weight tensors in state-of-the-art LLMs guarantee that there are sufficient tiles to concurrently utilize all available PIM units.

\noindent\textbf{Avoiding Cross-SIMD Compute.}
As discussed in \ssecref{jitq-pim-routine}, to compute the exponents on the MX block, the individual exponents of the elements are compared. 
With only \textit{pim-jitq-tiled}, elements of the same MX block (or subset of) are mapped to same DRAM word, resulting in cross-lane computations~\circled{D}. 
To avoid that, we propose to stride the tile across DRAM words, mapping each element to the same lane in each DRAM word~\circled{E}.
This ensures that elements of the same tile are aligned.
To eliminate the memory waste due to utilizing a single SIMD lane, we pack elements from independent tiles in the same DRAM word~\circled{E}. 
We term this mapping as \textit{pim-jitq-strided}.
As discussed above, the tensors from the evaluated LLMs have enough tiles to fully pack the SIMD lanes.

Note that, our proposed weight tensor placement spreads weight tensor across memory channels/banks and as such, as in baseline, can exploit memory parallelism effectively.

%% file: 06-evaluation.tex
\putsec{s06}{Evaluation}

\putssec{methodology}{Methodology}

\putsssec{eval_models}{Performance Models}

We analyze performance using analytical models as PIM is currently only available as part of functional prototypes ~\cite{samsungPIM}. 
Additionally, we aim to study highly optimized GPU implementations for the evaluated LLMs to provision a stronger GPU baseline to show PIM benefits over.
This makes relying on GPU simulators difficult and lends well to analytical models.

\noindent\textbf{GPU Performance Model.}
Our baseline GPU performance is assumed to be $max$(GPU compute time, GPU memory time), where GPU compute time considers GEMM operations (multiplies and adds) with peak compute throughput, while memory time considers only reading GEMM 's input tensors from HBM with 90\% of peak memory bandwidth.
In other words, we assume executing the vector operations in the transformer block, as well as writing the output of the transformer block's computation, to be free.
This is because, in optimized implementations, we observe that the non-GEMM (element-wise) operations are increasingly getting fused with the GEMMs, to avoid kernel launch and global memory access overheads. 
Additionally, to have an even stronger GPU baseline, we assume zero-overhead concurrent execution of weight and input gradient computations in the backward propagation phase of training which results in shorter LLM execution time on GPU.

\noindent\textbf{PIM Performance Model.}
We assume a PIM architecture in which the GPU issues PIM commands as special load/store accesses which bypass the caches and are issued in-order by the memory controller to multiple banks in parallel~\cite{samsungPIM}.
We take a detailed DRAM command orchestration approach in which, for a given weight matrix, we consider necessary data mapping (\ssecref{jitq-pim-data-placement}) and orchestration (\ssecref{jitq-pim-routine}). 
Next, we deduce the exact DRAM commands needed to orchestrate the computation.
We augment a detailed DRAM model for modeling PIM instruction timing that incorporates the PIM DRAM timing restrictions, including row activation overheads. 
We assume the parameters listed in \tabref{tab:pim_params}. 
Note that we assume a PIM-aware GPU which can issue \textit{pim-instructions} and \textit{pim-commands} at issue-rate. 
With the available thread parallelism at the GPU, we believe this to be a reasonable assumption. 

\input{table_pim_params}

\putsssec{eval_llms}{Evaluated LLMs} 
In this work, we evaluate LLMs of different hyperparameter combinations and distributed setups for training.
Specifically, we evaluate behaviour of models similar to the following LLMs ({model name}-{parameter count}): bert-345M~\cite{devlin2019bert}, gpt-2-1.5B~\cite{gpt2}, mega-lm-8.3B~\cite{shoeybi2020megatronlm}, t-nlg-17B~\cite{tnlg}, gpt-3-175B~\cite{gpt3}, mega-nlg-530B~\cite{smith2022-meganlg}, and palm-540B~\cite{chowdhery2022palm}.
Additionally, given the trend of how LLMs evolve, we project three future LLMs of sizes 1 trillion (future-1T), 10 trillion (future-10T), and 100 trillion (future-100T) parameters in \tabref{tab:eval_llms}. 
For these future models, we scale the number of layers, hidden dimension, and sequence length. 

\input{table_eval_future_llms}

\begin{figure}[t]
    \centering
    \includegraphics[width=\linewidth]{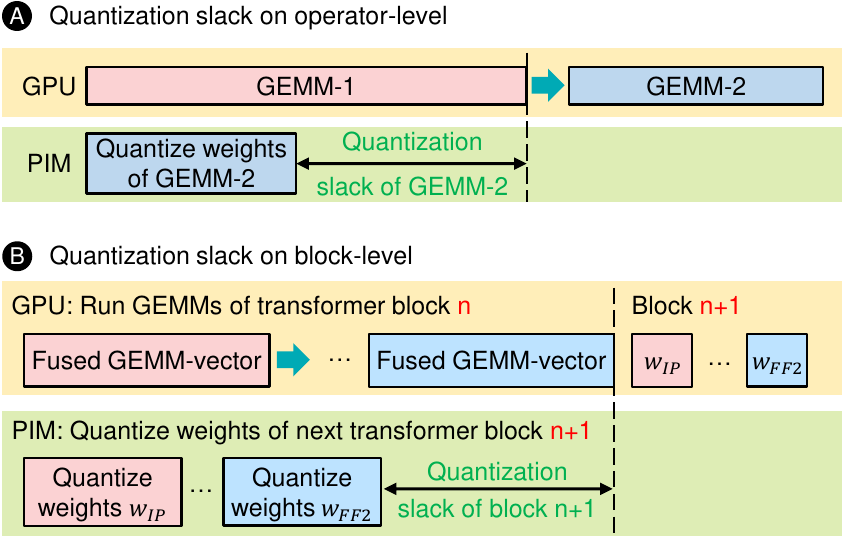}
    \caption{Quantization slack indicates PIM ability to perform JIT-Q without stalling concurrent GPU computation.}
    \label{fig:jit_q_slack}
    \vspace{-\baselineskip}
\end{figure}

\putsssec{eval_metrics}{Evaluated Metrics}
\noindent\textbf{JIT-Q Slack.}
To showcase the efficacy of offloading JIT-Q to PIM, we evaluate JIT-Q \textit{slack} as shown in \figref{fig:jit_q_slack}.
We define JIT-Q slack as the difference between the execution of GEMM $i$ on GPU, and the quantization time of the weights of the next GEMM $i$+1 on PIM~\circled{A}. 
Bigger slack indicates faster quantization on PIM.
Throughout this section, for simplicity, we model the next transformer block quantization instead of next GEMM as shown in~\circled{B}. 
That said, JIT-Q at the operator level should at least retain the capacity savings gained at the transformer block level.

\noindent\textbf{Memory Capacity Savings.}
We evaluate the capacity savings unlocked by JIT-Q on PIM due to storing a single temporary copy of the quantized weights of the next transformer block. 
For that, we estimate the capacity consumed by weights, gradients, activations, and optimizer state based on the following.
We assume a mixed precision training setup similar to FP8-based training~\cite{wang2018training,mellempudi2019mixed,graphcorefp16master} with BF16 high-precision master weights and optimizer state, MX$n$ low-precision weights, MX$n$ activations for GEMMs, and FP8 otherwise.
Additionally, for weights, we assume two copies for row and column quantization.
Finally, for activations, we assume a selective recomputation strategy~\cite{korthikanti2022reducing}. We also discuss implications with other mixed precision setups (e.g., FP32 master weights) in \ssecref{fp32_eval}. 

\begin{figure}[t]
    \centering
    \includegraphics[width=\linewidth]{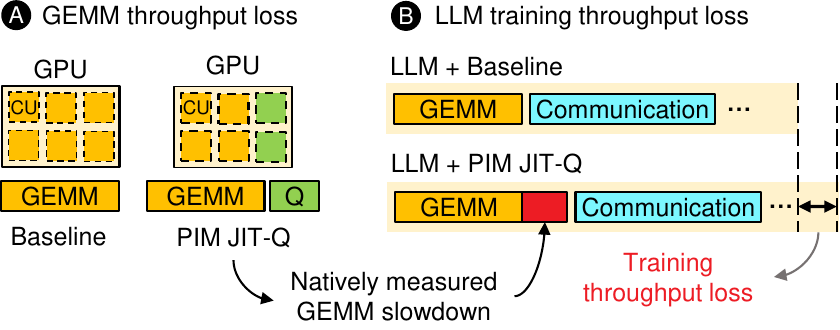}
    \caption{Compute interference between GPU and PIM execution can negatively impact training throughput.}
    \label{fig:jit_q_train_throughput_est}
    \vspace{-1.5\baselineskip}
\end{figure}

\noindent\textbf{Training Throughput Loss.}
As PIM quantization kernel is orchestrated by the GPU, running other training computation on GPU concurrently will require GPU compute resources (compute units or CUs or GPU cores) to be shared amongst the two as shown in \figref{fig:jit_q_train_throughput_est}. This loss of compute resources, along with contention for shared HBM resource can slowdown the concurrent GPU computation~\circled{A}. 
This in turn, can lead to training throughput loss~\circled{B}. 

To evaluate this loss in LLMs training throughput, we natively measure how the interference between GPU and PIM execution can slowdown the execution on GPU. Our system setup consists of an AMD Instinct\textsuperscript{\texttrademark} MI210 Accelerator comprising GPU with 104 CUs and four stacks of HBM2E memory for a total capacity of 64GB and a peak memory bandwidth of 1638.4 GB/s~\cite{mi210}. 
First, we use Omniperf~\cite{omniperf}, a system performance profiling tool for machine learning/HPC workloads running on AMD MI GPUs, to measure the slowdown of GPU compute while disabling 16 out of available 104 CUs. 
Our experiments show that a single CU can sustain necessary PIM command issue rate for two physical HBM channels. 
As such, with 32 channels in our system, we set aside 16 CUs to orchestrate PIM kernel.
Note that software and hardware optimizations can lower the CUs needed for PIM orchestration but we make a conservative assumption of 16 CUs to study worst-case resource requirement for PIM.
Second, we scale our modeled GPU compute time by the measured slowdown. Specifically, the execution on GPUs observes a slowdown only while concurrently running with PIM quantization. For example, with a large JIT-Q slack, the GPU and PIM execution interfere for a small portion of the execution, in which the GPU computations observe slowdown. Third, we estimate the LLM training time and its breakdown using an analytical performance model for LLMs as proposed in~\cite{pati2023computation}. Finally, with the updated GPU execution slowdown, we estimate the training throughput loss.

\putssec{pim_map_eval}{Evaluating PIM Mapping \& Orchestration}

\begin{figure}[t]
    \centering
    \includegraphics[width=\linewidth]{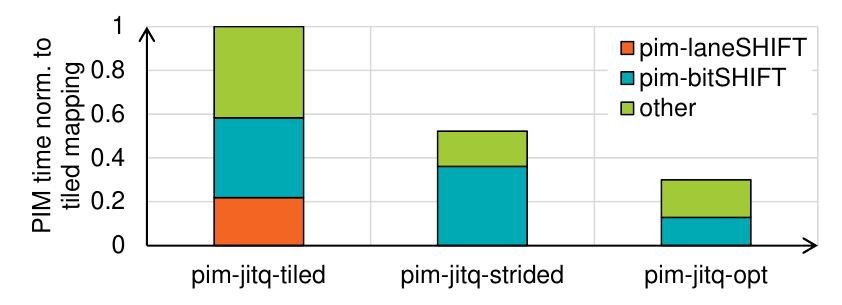}
    \caption{PIM row quantization time of the different weight tensor placements and optimizations (for BF16 to MX6).}
    \label{fig:results_pim_perf}
    \vspace{-1.5\baselineskip}
\end{figure}

In this section, we evaluate the performance of our proposed PIM JIT-Q mapping and orchestration.
Specifically, we compare the tiled mapping (\textit{pim-jitq-tiled} in \ssecref{jitq-pim-data-placement}), strided tiled mapping (\textit{pim-jitq-strided} in \ssecref{jitq-pim-data-placement}), and strided mapping with the optimization in \ssecref{jitq-pim-routine} (\textit{pim-jitq-opt}).
\figref{fig:results_pim_perf} depicts the PIM quantization time of of these PIM flavors normalized to baseline tiled mapping (\textit{pim-jitq-tiled}) on the y-axis for a representative LLM. 
Further, we break down the PIM quantization time into that spent executing shift-related PIM commands (\textit{pim-laneSHIFT} and \textit{pim-bitSHIFT}), and rest of the time as \textit{other} (contains DRAM row-open overhead, data movement from register to row-buffer, compare, etc.).
We observe that \textit{pim-jitq-strided} is superior to \textit{pim-jitq-tiled} (48\% lower quantization time) because of avoiding lane-level shifts (i.e., \textit{pim-laneSHIFT} commands).
Also, using \textit{pim-jitq-strided} not only eliminates \textit{pim-laneSHIFT}, but reduces other PIM commands as well. 
For example, by packing multiple tiles with strided mapping, fewer \textit{pim-CMP} commands are used to process the packed tiles compared to unpacked mapping.
Also, utilizing the augmented PIM design \textit{pim-jitq-opt} (\ssecref{jitq-pim-routine}) further reduces PIM quantization time to 70\% lower compared to \textit{pim-jitq-tiled} as it reduces the number of commands required for bit-level shifting to a single PIM command.  

\putssec{slack_eval}{PIM Quantizes without Stalling GPU}

\begin{figure}[t]
    \centering
    \includegraphics[width=\linewidth]{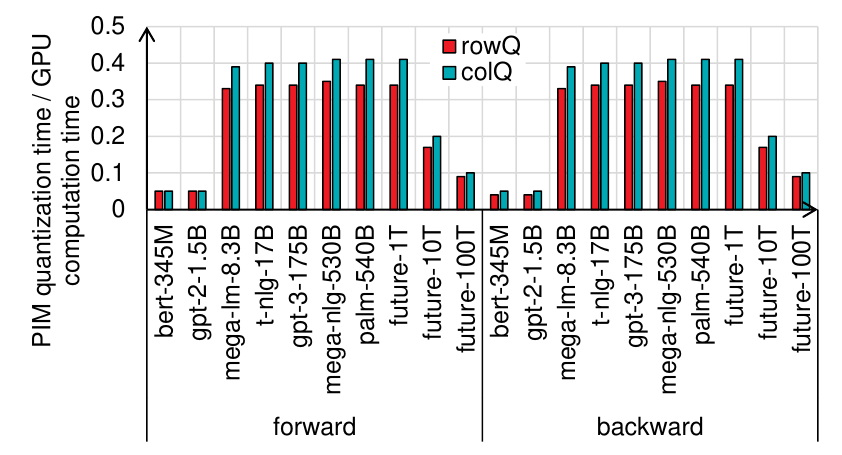}
    \caption{PIM JIT-Q slack (for BF16 to MX6).}
    \label{fig:results_jit_q_slack}
    \vspace{-\baselineskip}
\end{figure}

Next, we evaluate if via offload to PIM, weights can be quantized without stalling the GPU execution using the JIT-Q slack metric (\sssecref{eval_metrics}). 
\figref{fig:results_jit_q_slack} shows the JIT-Q slack for BF16 to MX6 as PIM quantization time normalized to GPU compute time (y-axis, lower is better) for the evaluated LLMs (x-axis). We observe that quantization with PIM exhibits sufficient slack for both forward and backward phases. Specifically, on average, we observe that the GPU computation time is at least 2.4$\times$ the quantization time with PIM. The large slack holds for narrower MX4 and wider MX9 formats with GPU time at least 1.6$\times$ and 3.7$\times$ PIM time, respectively (not shown).

We also observe that the varying slack across the evaluated LLMs is due to their sequence length (SL) and batch-size (B), which affect the quantization work on PIM and the compute work on GPU differently.
Specifically, increasing SL or B, as in bert-340M, gpt-2-1.5B, or models larger than future-1T, increases the input tensor size, but does not impact weight tensors. Thus GEMM compute increases, while quantization time on weights remains the same.

Additionally, we evaluate the performance of row and column quantization. 
We observe row quantization to have better slack compared to column quantization under forward and backward phases. 
This is due to the tiled data mapping discussed in \ssecref{jitq-pim-data-placement}. 
Specifically, with the row-major mapping of the tile in a bank, column elements end up in multiple DRAM rows resulting in additional DRAM rows opens for column quantization. 
This results in 18\% increase in column quantization time compared to row quantization. 

\putssec{capacity_eval}{Capacity Savings of PIM JIT-Q}

\begin{figure}[t]
    \centering
    \includegraphics[width=\linewidth]{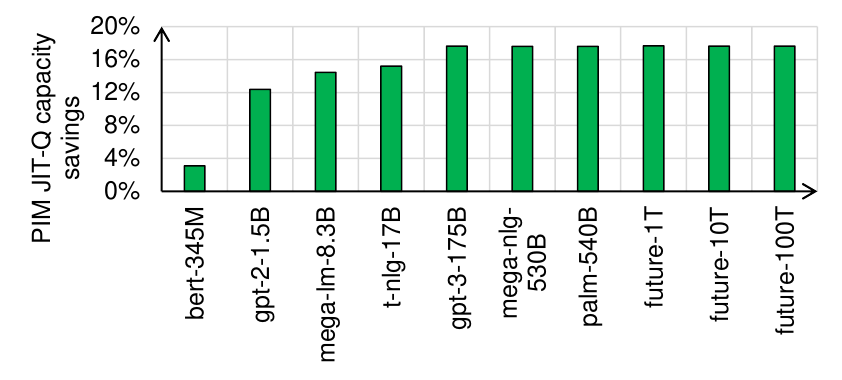}
    \caption{PIM JIT-Q capacity savings (for BF16 to MX6).}
    \label{fig:results_jit_q_capacity}
    \vspace{-1.5\baselineskip}
\end{figure}

\figref{fig:results_jit_q_capacity} shows capacity savings for BF16 to MX6. 
We observe a significant capacity savings of 12.5\%, on average, which continues to hold for future LLMs with trillions of parameters (up to 17.6\%). 
For narrow MX4 (not shown), the average capacity savings drops to 9\% (up to 12.5\%) as the overall capacity overheads of the low-precision weights gets smaller. 
In contrast, for wider MX9 (not shown), the average capacity savings increase to 16.8\% (up to 24.2\%).

The capacity savings unlock multiple benefits (not in tandem) such as training larger LLMs with the same available resources. 
For example, with MX6, it enables training a 20\% larger model compared to gpt-3-175B. 
Also, given that tensor slicing degree is largely dictated by underlying memory capacity, the capacity savings can lower the tensor-slicing degree needed.
This can lower the overall training cost, as well as potentially improving computation efficiency because of running larger GEMMs (less tensor slicing) per GPU, and reducing model-parallel related communication.
For example, the saved capacity enables the use of 12.5\% reduction in the tensor-slicing degree.
Further, the savings can allow larger batch-sizes to be used which in turn can improve computation efficiency and can lead to faster convergence. 
Finally, with the capacity savings, more activations can be stored which decreases the frequency of forward phase re-computation.

Finally, we also evaluate a baseline that maintains a single copy of the low-precision weights, either row or column, and observe that JIT-Q delivers, on average, capacity savings of 4.8\% (up to 6.6\%), 6.7\% (up to 9.7\%), and 9.3\% (up to 13.8\%) for MX4, MX6, and MX9, respectively, averaged across the evaluated LLMs.
Note that such baseline would incur higher data movement due to reading the high-precision copy to generate the required low-precision copy.

\putssec{throughput_eval}{Effect of PIM JIT-Q on Training Throughput}

\begin{figure}[t]
    \centering
    \includegraphics[width=\linewidth]{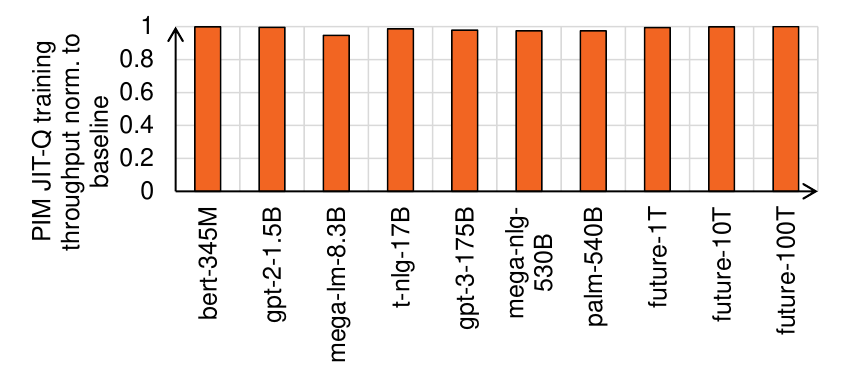}
    \caption{PIM JIT-Q training throughput loss (for BF16 to MX6).}
    \label{fig:results_jit_q_train_throughput}
    \vspace{-1.5\baselineskip}
\end{figure}

Assuming the methodology in \sssecref{eval_metrics}, we estimate the training throughput loss shown in \figref{fig:results_jit_q_train_throughput}. 
We observe, for BF16 to MX6 quantization that PIM delivers capacity savings at small training throughput loss of 1.6\% on average (up to 5\% for mega-lm-8.3B). 
For BF16 to MX4 (not shown), the average throughput loss is 2.4\% due to the increased overlap between GPU and PIM. 
In contrast, the average throughput loss for BF16 to MX9 is 1.1\% (not shown).
It is worth noting that the reduction in forward phase re-computation enabled by the capacity savings can positively affect training throughput and possibly regain/improve throughput loss due to interference. 
Further, larger batch-sizes, enabled by the capacity savings due to JIT-Q, can also help recover this loss. 
Finally, in future PIM designs that offloads PIM orchestration to dedicated engines instead of GPU, this compute interference will not pose a challenge.

\putssec{fp32_eval}{PIM JIT-Q with FP32 Master Weights}
The single bit shifts to deduce the output MX mantissa bits (\textit{pim-bitSHIFT}) are proportional to the number of mantissa bits in the input scalar format.  
Therefore, for FP32 master weights, the bit-level shift commands increase, which increases PIM quantization time, and subsequently affects the quantization slack.  
Specifically, on average, we observe that the GPU compute time is at most 0.51$\times$ and 0.76$\times$ of the PIM quantization time for MX4 and MX6, respectively, while maintaining the slack for MX9.
That said, although PIM is not able to quantize the weights for the next transformer block $i$+1, PIM JIT-Q can prepare for $i$+2 instead.
In other words, we maintain the quantized weights for the next two transformer blocks instead of one. 
Compared to keeping the low-precision copies for all the weight tensors in the baseline, JIT-Q delivers  capacity savings of 5.6\%, 7.9\%, and 11.1\% at throughput loss of 4.4\%, 4.2\%, and 3.6\% for MX4, MX6, and MX9, respectively, averaged across the evaluated LLMs. 

%% file: table_pim_params.tex
\begin{table}[t]
\caption{Parameters for performance model~\cite{HBM3-jedec}.}
\label{tab:pim_params}
\vskip 0.15in
\begin{center}
\begin{small}
\begin{tabular}{l l}
\toprule
\#Banks per Stack (4-high)      & 512 \\ 
\midrule
Bandwidth per Pin               & 4.8 Gb/s \\              
\begin{tabular}[c]{@{}l@{}}GPU Memory Bandwidth\\ per Stack\end{tabular} & 614.4 GB/s \\
\midrule
Row Buffer Size                 & 1024 B \\
\midrule
DRAM Parameters                 & \begin{tabular}[c]{@{}l@{}}tRP = 15ns, tCCDL=3.33ns,\\ tRAS=33ns\end{tabular} \\
\midrule
PIM Parameters & \begin{tabular}[c]{@{}l@{}}\#PIM Units per Stack = 256\\ \#PIM Registers per ALU = 16\end{tabular} \\ 
\bottomrule
\end{tabular}
\end{small}
\end{center}
\vskip -0.1in
\vspace{-\baselineskip}
\end{table}

%% file: table_eval_future_llms.tex
\begin{table}[t]
\caption{Future LLMs. L = \#layers, H = hidden dimension, A = \#attention heads, P = \#parameters, SL = sequence length, B = batch size, TP/PP = tensor/pipeline parallelism degree.}
\label{tab:eval_llms}
\vskip 0.15in
\begin{center}
\begin{small}
\begin{tabular}{cccccccc}
\toprule
LLM&                        L&   H&      A&   SL&    B&  TP&   PP\\
\midrule
future-1T&                  80&  32K&  128& 4K&  1&   128& 2\\
future-10T&                 200& 64K&  128& 8K&  1&   128& 8\\
future-100T&                500& 128K& 128& 16K& 1&   128& 32\\
\bottomrule
\end{tabular}
\end{small}
\end{center}
\vskip -0.1in
\vspace{-\baselineskip}
\end{table}

%% file: 07-discussion.tex
\putsec{s07}{Discussion}

\noindent\textbf{Master Weights Sharding.} 
Works such as Zero-Redundancy parallelism (ZeRO)~\cite{samyam2020zero} and FSDP~\cite{zhao2023pytorch} shards the model states (parameters, gradients, and optimizer state) across GPUs instead of replicating them.
The required tensor is reconstructed (communicated) on-demand before computations.
Our proposed JIT-Q can be utilized per each (PIM-enabled) GPU to locally quantize its shard of the master weights before communicating the parameters. 
That said, for directional blocked formats (MX), the partitioning of master weight tensor should be in tiled fashion (\ssecref{jitq-pim-data-placement}).
Finally, JIT-Q can also be used for or in conjunction with recent works that further quantize weights (and gradients) to reduce overall communication volume~\cite{wang2023zeroplusplus}. 

\noindent\textbf{Master Weights Placement.}
In scenarios where master weights are offloaded to CPU memory, and only the low-precision copy is sent to/resides in GPU~\cite{ren2021zerooffload,ZeRO-Infinity21}, JIT-Q can be done on PIM-enabled CPU memory to eliminate the persistent low-precision weight copies in GPU memory. In this case, JIT-Q only changes when to send the low precision weights from CPU to GPU, not the CPU-GPU traffic volume.

\noindent\textbf{Blocked Formats Variants.}
We discuss MX formats as exemplar of blocked data formats.
That said, PIM routines for other block sizes with more/fewer exponent levels~\cite{rouhani2023microscaling} and different sub-block granularity can be deduced and employed.

\noindent\textbf{Scalar-to-scalar JIT-Q.}
Quantization from one scalar format to another (e.g., from FP32 to BF16), used in existing training setups, is simpler compared to MX quantization. 
This is because each input is independent (no blocking) making the computation element-wise, thus even more PIM-amenable. 
Also, it does not involve computing the shared exponents, and the bit shift amount is the same across all the inputs.
Therefore, we can simplify our PIM quantization routine and mapping to perform such quantization. 

%% file: 08-related.tex
\putsec{s08}{Related Work}

Several ML techniques are employed to reduce per-device memory requirements. Distributed techniques help increase the effective memory capacity by slicing model parameters across multiple devices; pipeline parallelism maps layers to devices~\cite{yanping2019_gpipe,deepak2019_pipedream}, tensor/sequence parallelism slice individual layers~\cite{narayanan2021efficient,li2022sequence}, and ZeRO/FSDP shard model states across data-parallel devices~\cite{samyam2020zero,zhao2023pytorch}. Other works offload model states to heterogeneous system memory (e.g., CPU, NVMe)~\cite{ren2021zerooffload,ZeRO-Infinity21}. Finally, activation checkpointing trades-off computation for capacity~\cite{chen2016training,korthikanti2022reducing}. JIT-Q reduces the reliance on such techniques and can also be used in conjunction to limit the required distributed device count, communication overheads, and extraneous compute.    

Similarly, compression can provide memory savings~\cite{nikolic2022schrodingers,zadeh2022_mokey} or reduce communication~\cite{wang2023zeroplusplus,youhui2021_gredcompress,zhen2022_mics,rhu2018_cdma}.
One class of compression is quantization which is used for both training and inference~\cite{micikevicius2022fp8,noune20228bit,Rouhani23MX}.
While we focus on MX formats, our proposal can be extended to other data formats. 

Additionally, ML algorithm-aware techniques such as model pruning~\cite{anwar2017structured}
and knowledge distillation~\cite{hinton2015distilling} reduce capacity needs by removing unnecessary model parameters. These however are employed during deployment and do not help with capacity savings during training. Furthermore, these techniques require algorithmic understanding and/or additional training to preserve model accuracy. 

Finally, many works leverage PIM's data movement and performance benefits to accelerate ML~\cite{samsungPIM,hynixPIM,pati2022_bert,oliveira2022_pimnn,aga2019_coml}.
To the best of our knowledge, this is the first work which showcases PIM's ability to limit capacity requirements via efficient JIT quantization.

%% file: 9-conclusions.tex
\putsec{s09}{Conclusion}

Memory capacity is a key determinator of ML efficiency. 
This work proposes harnessing of processing-in-memory for just-in-time quantization of weight tensors. 
This eliminates weight tensor redundancy to deliver memory capacity savings. 
We show that our proposed PIM quantization routine can keep up with GPU computation and deliver up to 24\% memory capacity savings at marginal throughput loss (up to 2.4\%) for LLM training. 
The capacity savings can be harnessed to train larger models, to reduce the number of GPUs required for training, or for better compute efficiency.